\documentclass[prd,preprint,superscriptaddress]{revtex4}

\begin{document}

\title{Anisotropy and inflation in Bianchi I brane worlds}

\author{Juan M. Aguirregabiria}
\affiliation{Fisika Teorikoa eta Zientziaren Historia Saila,
Zientzi Fakultatea,
Euskal Herriko Unibertsitatea,
644 Posta Kutxatila, 48080 Bilbao, Spain}
\author{Luis P. Chimento}
\affiliation{Departamento de F\'{\i}sica,
Facultad de Ciencias Exactas y Naturales,
Universidad de Buenos Aires,
Ciudad  Universitaria,  Pabell\'on  I,
1428 Buenos Aires, Argentina.}
\author{Ruth Lazkoz}
\affiliation{Fisika Teorikoa eta Zientziaren Historia Saila,
Zientzi Fakultatea,
Euskal Herriko Unibertsitatea,
644 Posta Kutxatila, 48080 Bilbao, Spain}

\begin{abstract} After a more general assumption on the influence of the
bulk on the  brane, we extend some conclusions by Maartens 
\emph{et al.}~\cite{Maartens} and Santos \emph{et al.}~\cite{Santos} on the
asymptotic behavior of Bianchi~I brane worlds. As a consequence of the
nonlocal anisotropic stresses induced by the bulk, in most of our
models, the brane does not isotropize and  the nonlocal energy does not
vanish in the limit in which the mean radius goes to infinity. We have
also found the intriguing possibility that the inflation due to the
cosmological constant might be prevented by the interaction with the
bulk. We show that the problem for the mean radius  can be completely
solved in our models, which include as particular cases those in the
references above. \end{abstract}

\date{\today}

\maketitle

%%%%%%%%%%%%%%%%%%%%%%%%
\section{Introduction}
\label{sec:intro}

The main stream approach to try and solve the problems arising from the
break down of Einstein's theory at high energies is that of considering
it a particular limit of a more general theory, as happened with
Newton's theory in relation to that of Einstein's. One of this schemes
is the brane gravity picture \cite{Horava,Randall}, according to which
matter fields are confined to a hypersurface (the brane) with three
spatial dimensions embedded in a higher dimensional space (the bulk) on
which gravity can act also.

Considerable efforts in this area of research have been directed to testing
that scheme by deducing cosmological implications from it. The usual
framework for these studies is the geometric approach of
\cite{Shiromizu}, which provides the gravitational field equations
induced on the brane along with conservation equations for bulk degrees
of freedom.  Reviews on this topic can be found, for instance, in
\cite{Maartens-rev,Langlois,Brax}.

Studies concerning FRW universes \cite{Binetruy} hinted that the
dynamics of the early Universe in this new scenario is rather peculiar
because of the modifications in the Friedmann equation. Then, when
anisotropic models were considered \cite{Campos}, it was shown that 
discrepancies with respect to the standard scenario extended to the
shear dynamics as well. 

In the set-up of general relativity, Wald \cite{Wald} showed that a positive
cosmological constant leads to the isotropization of expanding
homogeneous cosmological models .  The anisotropy dissipation
in brane world inflation in the absence of effective cosmological
constant was analyzed by Maartens, Sahni and Saini \cite{Maartens}. 
Taking into account the cosmological constant,  Santos, Vernizzi and
Ferreira derived a set of sufficient conditions that allow extending
Wald's result to the brane world scenario \cite{Santos} (see also
\cite{Paul} for an alternative view on the same problem). The last three
works were done directly on the brane, where an additional hypothesis
has to be done for the nonlocal anisotropic stresses induced by the
bulk, for they are not given by any evolution equation on the brane.
This naturally raises the question of to what extent the results depend on
the aforementioned hypothesis.

The goal of this work is to extend the results in \cite{Maartens,Santos} by
considering a more general additional hypothesis. We will analyse, as
done in those works, a Bianchi I brane embedded in a 5-dimensional bulk.
We assume that the nonlocal stresses satisfy a condition which reduces
to the one considered in Refs.~\cite{Maartens,Santos} in particular
cases. We will show that the evolution equations on the brane can be
integrated  and conclude that, generically, we have exponential inflation
or an asymptotic power law for the mean radius,  but the models do not
isotropize unless they belong to the class considered in the
aforementioned works.

%%%%%%%%%%%%%%%%%%%%%%%%
\section{Bianchi I brane models}
\label{sec:equations}

Let us consider a Bianchi I brane with the induced metric
 \begin{equation}\label{eq:metric}
 ds^2=-dt^2+a_1^2(t)\,dx^2+a_2^2(t)\,dy^2+a_3^2(t)\,dz^2
 \end{equation}
when the matter on the brane is a perfect fluid
of density $\rho$ and pressure $p$. We define as usual
the mean radius $a\equiv(a_1a_2a_3)^{1/3}$ and 
the mean Hubble parameter $H\equiv\dot a/a=\frac13 \sum_i H_i$, with $H_i=\dot a_i/a_i$.  

We will be using the equations induced on the brane derived by
Shiromizu, Maeda and Sasaki~\cite{Shiromizu} but follow the notation of \cite{Maartens,Santos}.
For the metric (\ref{eq:metric}), the dynamics of the mean radius $a$ in
this model can be described by the following equations on the brane:
\begin{eqnarray}
3H^2&=&
\frac{6}{\kappa^2\lambda}\mathcal{U}+\frac12\sigma^2+\Lambda+\kappa^2\rho\left(1+\frac\rho{2\lambda}\right),\label{eq:hsquare}\\
  3\dot H+3H^2+\sigma^2+\frac{6}{\kappa^2\lambda}\mathcal{U} &=&\Lambda-\frac{\kappa^2}2(\rho+3p)-\frac{\kappa^2}2(2\rho+3p)\frac\rho\lambda,\label{eq:doth}\\
\dot\sigma_{\mu\nu}+3H\sigma_{\mu\nu}&=&\frac{6}{\kappa^2\lambda}\mathcal{P}_{\mu\nu},\label{eq:dotsigma}\\
\dot\mathcal{U}+4H\mathcal{U}+\sigma^{\mu\nu}\mathcal{P}_{\mu\nu}&=&0,\label{eq:dotu},
\end{eqnarray}
where a dot denotes $u^\mu\nabla_\mu$, with $u^\mu =  \partial/\partial t$. Throughout the paper $\kappa^2/8\pi$ and $\tilde\kappa^2/8\pi$ will, respectively, be the effective Newton constant on the brane and in the bulk ;
$\Lambda$ will be the effective 4D cosmological constant, and the tension on the brane will be $\lambda\equiv
6\kappa^2/\tilde\kappa^4$. On the other hand, $\mathcal{U}$ stands for the effective nonlocal energy density, and $\sigma_{\mu\nu}$ denotes the shear scalar, which satisfies (\ref{eq:dotsigma}) and
 \begin{equation}
 \sigma^2\equiv\sigma^{\mu\nu}\sigma_{\mu\nu} = \sum_{i=1}^3{\left(H_i-H\right)^2}.
 \end{equation}
In addition, if $u^{\mu}$ is  the four
velocity of an observer on the brane comoving with matter, then
$h_{\mu\nu}=g_{\mu\nu}+u_{\mu}u_{\nu}$ projects into the comoving
rest-space, and $\cal U$ and ${\cal P}_{\mu\nu}$ will be related to 
${\cal E}_{\mu\nu}$, which is the projection on the brane of the 5D Weyl tensor,
through
\begin{eqnarray}
{\cal U}&=&-\frac{\kappa^2\lambda}{6}{\cal E}_{\mu\nu}u^{\mu}u^{\nu},\\
{\cal P}_{\mu\nu}&=&-\frac{\kappa^2\lambda}{6}\left(h_{\mu}^{\;\alpha}h_{\nu}^{\;\beta}-\frac{1}{2}
h^{\alpha \beta}h_{\mu\nu}\right){\cal E}_{\alpha \beta}.
\end{eqnarray}

There is also, in principle, the constraint that the
projected spatial covariant derivative of $\mathcal{P}_{\mu\nu}$ vanishes:
 \begin{equation}\label{eq:constraint}
 \mathrm{D}^\nu\mathcal{P}_{\mu\nu}=0,
 \end{equation}
 but this result holds identically for the metric (\ref{eq:metric}). 
On the other hand, as a consequence of system (\ref{eq:hsquare})--(\ref{eq:dotu}), we get
the conservation law
 \begin{equation}
\dot\rho+3H(\rho+p)=0.\label{eq:dotrho}
 \end{equation}

%%%%%%%%%%%%%%%%%%%%%%%%
\section{Exact examples and asymptotic behavior}
\label{sec:vacuum}

Since there is no evolution equation on the brane for the nonlocal
anisotropic stress $\mathcal{P}_{\mu\nu}$,  but only the constraint
(\ref{eq:constraint}), which in this case does not provide any information, some additional hypothesis is necessary 
to integrate the system (\ref{eq:hsquare})--(\ref{eq:dotu}).

In order to get some insight on the problem, we will restrict ourselves
for a moment to the vacuum case $\rho=p=0$, where  we get from
(\ref{eq:hsquare})--(\ref{eq:doth})
 \begin{equation}\label{eq:uvacuum}
 \mathcal{U}=\frac{\kappa^2\lambda}2\left(\dot H+3H^2-\Lambda\right).
 \end{equation}
Let us consider under which conditions the stable asymptotic behavior
may approach a power law $a\sim t^k$, ($k>0$). In such a case $\dot H, H\to0$, 
$\mathcal{U}\to-\kappa^2\lambda\Lambda/2$ and, because of
(\ref{eq:dotu}),
 \begin{equation}\label{eq:limsp}
 \sigma^{\mu\nu}\mathcal{P}_{\mu\nu}\to 2 \kappa^2\lambda\Lambda H.
 \end{equation}
We will show in the following that, if one chooses
the unknown $\mathcal{P}_{\mu\nu}$ so that this condition is satisfied,
 the asymptotic behavior described by a power law is stable. 
Furthermore, this will happen even in the presence of a fluid.

%%%%%%%%%%%%%%%%%%%%%%%%
%\section{The main assumption}
\label{sec:assumption}

So as to integrate the system (\ref{eq:hsquare})--(\ref{eq:dotu}) one
often \cite{Maartens} assumes $\mathcal U=0$ or the more general
$\sigma^{\mu\nu}\mathcal{P}_{\mu\nu}=0$, which has been used in
\cite{Santos} to discuss the stability of the de Sitter spacetime. It has been proved \cite{Hoogen} that spatial homogeneity follows from the integrability conditions for vanishing 
non-local anisotropic stress and energy flux, which are two of the three bulk degrees of freedom. Therefore, the considerable simplification arising from switching off those quantities is consistent with having a
Bianchi I metric on the brane.  

It must be pointed out that other choices can be found in the literature;  Barrow and Maartens \cite{Barrow}, in an investigation of early times shear anisotropy in an inhomogeneous universe,
assumed that ${\cal P}_{\mu\nu}$ behaves qualitatively like a general 4D anisotropic stress, in particular they chose ${\cal P}_{\mu\nu}$ to be proportional to the energy density of
the anisotropic source (which we will denote with $\tilde\rho$) so that 
${\cal P}_{\mu\nu}=\tilde\rho \,C_{\mu\nu}$ with $\dot C_{\mu\nu}=0$, $\sqrt{C_{\mu\nu}C^{\mu\nu}}={\cal O}(1)$, 
and $\tilde\rho\ll\rho$.

Although the asymptotic behavior (\ref{eq:limsp}) may happen with many
choices of  $\sigma^{\mu\nu}\mathcal{P}_{\mu\nu}$, we will consider only
a simple family of models.  In the following, we will assume that
$\sigma^{\mu\nu}\mathcal{P}_{\mu\nu}$ is proportional to the Hubble
parameter, so that it can be written as
 \begin{equation}
 \sigma^{\mu\nu}\mathcal{P}_{\mu\nu}=2\kappa^2\lambda \Gamma(a)\,H,\label{eq:assumption}
 \end{equation}
which reduces to the cases discussed in~\cite{Maartens,Santos} when the
function $\Gamma(a)$ vanishes. By making this hypothesis, we use a less
restrictive assumption to check whether the conclusions
reached in \cite{Maartens,Santos} hold in a more general context, while
still being able to integrate the evolution equations. Note that 
(\ref{eq:limsp}) is recovered, even before the asymptotic behavior is
reached, provided $\Gamma(a)=\Lambda$. 

Exactly as happens with the restrictive $\Gamma=0$ case often used in
the bibliography, it remains an open problem whether the more general
assumption we make is compatible with a full 5D
solution (see \cite{Matravers} for a recent attempt at tackling the 5D
problem). Judging on the suitability of such guesses is, therefore,
not possible for the time being, but they definitely help unveiling
possible unexpected consequences of the interference between bulk and
brane.

%%%%%%%%%%%%%%%%%%%%%%%%
%\section{Solving the equations}
\label{sec:solving}

The nonlocal energy density $\mathcal{U}$ is given by the conservation
equation (\ref{eq:dotu}), which is easily solved under the assumption
(\ref{eq:assumption}):
 \begin{equation}\label{eq:uint}
  \mathcal{U}=\frac{\kappa^2\lambda u_0}{a^4}-\frac{2\kappa^2\lambda}{a^4}\int a^3\Gamma(a)\,da,
 \end{equation}
where $u_0$ is an arbitrary integration constant.
We see here that,  for expanding universes, 
\begin{equation}\label{eq:uinta}
 \lim_{a\to\infty}\mathcal{U}= -\frac12\kappa^2\lambda \lim_{a\to\infty}\Gamma(a), 
\end{equation}
(provided the limit on the right side exists or is infinite)
so that it  does not always vanish asymptotically, although it will 
go to zero in the special cases
in which   $\Gamma(a)\to0$ (which of course include the choice
$\Gamma(a)=0$ of Refs.~\cite{Maartens,Santos}), where it reduces to the
result  by Toporensky \cite{Toporensky}. Expression (\ref{eq:uinta})
suggests there is the possibility that,   as a consequence of the
interaction of the brane with the bulk through $\mathcal{P}_{\mu\nu}$,
the asymptotic vanishing of $\mathcal{U}$ arising in the cases studied
in \cite{Santos} might be evitable.

By contracting (\ref{eq:dotsigma}) with $\sigma^{\mu\nu}$ and using
(\ref{eq:assumption}), one readily gets
 \begin{equation}\label{eq:sigmaint}
 \sigma^2=\frac{\sigma_0^2}{a^6}+\frac{24}{a^6}\int a^5\Gamma(a)\,da,
 \end{equation}
for any constant $\sigma_0$, so that
 \begin{equation}\label{eq:sigmaint1}
 \lim_{a\to\infty}\sigma^2= 4\lim_{a\to\infty}\Gamma(a).
 \end{equation}
Then, it is clear that asymptotically $\Gamma$ must be non-negative. 
If $\Gamma\to0$ the brane world isotropizes as $a\to\infty$, as
described in \cite{Maartens,Santos}, but, for all other asymptotic behaviours 
of $\Gamma$, there will be a remaining anisotropy
induced by the nonlocal anisotropic stresses. Note that there is also the possibility
that the anisotropy grows with $a$, which would correspond to ${\cal U}<0$ at late
times. Models which do not isotropize in a recollapsing situation, unlike ours, were found in \cite{Sopuerta}, and they were characterized by ${\cal P}_{\mu\nu}$  and ${\cal U}<0$.

Clearly, our work and many others suggest that the  interplay between the bulk degrees of freedom and the dynamics on the brane is rather non-trivial. An interesting illustration of this is a recent work \cite{Savchenko} which investigated the dynamics of a flat isotropic brane world with a perfect fluid with equation of state 
$p=(\gamma-1)\rho$ and a scalar field with a power-law potential. There it was found
that the number  and the stability of fixed points of the system describing the dynamics 
of the model would depend not only on  whether ${\cal U}$ vanishes or  not, but also on 
its sign.

If we assume the equation of state $p=(\gamma-1)\rho$, the conservation law (\ref{eq:dotrho})
is equivalent to
 \begin{equation}\label{eq:rhoint}
 \rho=\frac{\rho_0}{\kappa^2a^{3\gamma}},
 \end{equation}
with an arbitrary constant $\rho_0$.

If we insert (\ref{eq:uint}), (\ref{eq:sigmaint}) and (\ref{eq:rhoint})
in (\ref{eq:hsquare}), we get the generalized Friedmann equation
 \begin{equation}\label{eq:Friedmann}
 3H^2=\frac{6u_0}{a^4}+\frac{\sigma_0^2}{2a^6}+
      \frac{\rho_0}{a^{3\gamma}}+\frac{\rho_0^2}{2\kappa^2\lambda a^{6\gamma}}+
      \Lambda-\frac{24}{a^6}\int\left[ a\int a^3\Gamma(a)\,da\right]\,da.
 \end{equation}

The orbits $(H(\mathcal{U}),\sigma(\mathcal U))$ were found
in~\cite{Santos} for the special case $\Gamma=0$, but we can see that,
for any $\Gamma(a)$, the problem for $a(t)$, $\sigma(t)$ and
$\mathcal{U}(t)$  may be completely solved from equation
(\ref{eq:Friedmann}) by means of a quadrature and a function inversion.
The integrals may be explicitly computed (in terms of elementary or
elliptic functions)  for different choices for the constants
$u_0$, $\sigma_0$, $\rho_0$, $\gamma$ and $\Lambda$ and the function
$\Gamma(a)$.  The simplest cases are solutions with no fluid
($\rho_0=0$). If $\Gamma(a)=\Lambda$,
%as
% \begin{equation}\label{eq:part}
% w\sqrt{1+w^2}-\log\left(w+\sqrt{1+w^2}\right)=\frac{18\sqrt{3u_0^3}}{\sigma_0^2}\,t,
% \qquad w\equiv\frac{3u_0^{1/2}}{\sigma_0}\,a.
% \end{equation}
by using the parameter $0<w<\infty$ one may write the solution as
 \begin{eqnarray} 
 t&=&\frac{\sigma_0^2}{48\sqrt{2u_0^3}}\left(\sinh w -w\right),\label{eq:part1}\\\label{eq:part2}
 a&=&\frac{\sigma_0}{2\sqrt{3u_0}}\,\sinh \frac{w}{2},
 \end{eqnarray}
which at late time  corresponds to $a\sim t^{1/2}$. 

In contrast, if $u_0=\rho_0=0$ and $\Gamma(a)=\alpha$ for a constant $\alpha<\Lambda$, 
we get
\begin{eqnarray}
a^6&=&\frac{\sigma_0^2}{2(\Lambda-\alpha)}\,\sinh^2\sqrt{3(\Lambda-\alpha)}\,t.
\end{eqnarray}

Other exact solutions with $u_0=0$ can obtained with dust ($\gamma=1$)
or a stiff fluid ($\gamma=2$). Let us look first at the dust cases,
which, arguably,  are very interesting from the observational point of
view \cite{Dabrowski}. For $\Gamma(a)=\alpha$ we have
 \begin{equation}\label{v_lambda_zero_u_zero}
a^3=\frac{\rho_0}{\Lambda -\alpha}\,\sinh^2 \frac{\sqrt{3(\Lambda -\alpha)}\,t}{2}+
  \sqrt{\frac{\rho_0^2 + \kappa^2\lambda\sigma_0^2}
  {2\kappa^2\lambda(\Lambda-\alpha)}}\,\sinh\sqrt{3(\Lambda-\alpha)}\,t.
\end{equation}
The latter is a generalization of the Heckmann-Shucking metric~\cite{Heckmann,Khalatnikov} that
has not been discussed in the literature so far. Clearly, the
$\alpha\to\Lambda$ limit of the solution   (\ref{v_lambda_zero_u_zero})
is regular, with the form of  a second-order polynomial in $t$. The effects of the shear and the quadratic corrections  are of the
same order, they dominate at early times, and $a\sim t^{1/3}$. In contrast, at late times the model is neither aware of the anisotropy
nor of the extra dimensional ingredients, and $\log a\propto t$, or  $a\sim t^{2/3}$ if $\Lambda=\alpha$. On the
other hand, for a stiff fluid ($\gamma=2$), and $\Gamma(a)=\alpha$, we
have
 \begin{equation}  
   a^6= \frac{2\,\rho_0+\sigma_0^2}{2(\Lambda-\alpha)}\,\sinh ^2{\sqrt{3(\Lambda-\alpha)}\,t}+ 
   \frac{\rho_0}{\sqrt{2\,\kappa^2\,\lambda(\Lambda -\alpha)}}\,
   \sinh 2{\sqrt{3( \Lambda-\alpha) }\,t}.
 \end{equation} 
This solution too has a regular 
$\alpha\to\Lambda$ limit in the form of  a second-order polynomial in $t$.
Like in the dust solution above, the effects of the shear and the quadratic corrections  
are equally important, but  they dominate at late times,  instead; in that regime  
we either have $\log a\propto t$, or $a\sim t^{1/3}$ if $\Lambda=\alpha$. On the
contrary, at early times the model becomes  isotropic and relativistic, and $a\sim t^{1/6}$.

We can also obtain directly from (\ref{eq:Friedmann}) some general late-time
results of interest.  If asymptotically $\Gamma(a)\sim\alpha$ is constant,
then for $\Lambda-\alpha>0$, the brane inflates exponentially when
$a\to\infty$, as happened for $\alpha=0$, $\Lambda>0$ \cite{Santos}. But now
there is another  interesting possibility: the repulsion due to the
cosmological constant may be neutralized by the bulk influence in which case
the asymptotic behavior of $a$ is a power law. For simplicity we show this in
vacuum and write (\ref{eq:uvacuum}) in terms of the variable $\mu=-2\dot
H/3H^2$, which becomes a constant $\mu_0$ for the power-law solutions
$a\propto t^{2/3\mu_0}$, so

\begin{equation}
\label{mu.}
\dot\mu+\left[3H\mu-\frac{8H\mathcal{U}+2\sigma^{\mu\nu}P_{\mu\nu}}
{2\mathcal{U}+\kappa^2\lambda\Lambda}\right](2-\mu)=0.
\end{equation}

\noindent The existence of the constant solution, $\mu=\mu_0$, in the last
equation requires that

\begin{equation}
\label{sp}
\sigma^{\mu\nu}P_{\mu\nu}=3H\left[\mu_0\left(\mathcal{U}+
\frac{\kappa^2\lambda\Lambda}{2}\right)-\frac{4}{3}\mathcal{U}\right],
\end{equation}

\noindent this means that, for any solution of (11) which asymptotically
approaches to a power law, then the quantity $\sigma^{\mu\nu}P_{\mu\nu}$
asymptotically behaves as (\ref{sp}). Now, inserting (\ref{sp}) in
(\ref{eq:dotu}) and (\ref{eq:dotsigma}), we get the asymptotic expression for
the effective nonlocal energy density

\begin{equation}
\label{u}
\mathcal{U}=\frac{\kappa^2\lambda \alpha }{a^{3\mu_0}}-
\frac12\kappa^2\lambda\Lambda,
\end{equation}
and the shear scalar

\begin{equation}
\label{s}
\sigma^2=\frac{\sigma_0^2}{a^6}+
\frac{12\alpha (\mu_0-4/3)}{(2-\mu_0)a^{3\mu_0}}+4\Lambda,
\end{equation}

\noindent where $\alpha $ and $\sigma_0^2$ are arbitrary constants. In addition,
introducing (\ref{sp}) into (\ref{mu.}), it reads as

\begin{equation}
\label{eq:mueq}
\dot\mu=-3H\left(\mu-\mu_0\right)(2-\mu).
\end{equation}

\noindent It is easy to see that, for ordinary fluids, $\mu_0<2$, the
solutions of Eq.~(\ref{eq:mueq}) converge, respectively, to the stable fixed
point $\mu=\mu_0$, which describes the asymptotic power-law solutions
$a=t^{2/3\mu_0}$. For instance, if we additionally assume a perfect fluid
source, after using (\ref{eq:rhoint}), (\ref{u}) and (\ref{s}), the Einstein
equation (\ref{eq:hsquare}) becomes

\begin{equation}
\label{00}
3H^2=\frac{6\alpha }{a^{3\mu_0}}+\frac{\sigma_0^2}{2a^6}+
\frac{4\alpha }{(2-\mu_0)a^{3\mu_0}}+
\frac{\rho_0}{a^{3\gamma}}+\frac{\rho_0^2}{2\kappa^2\lambda a^{6\gamma}}.
\end{equation}

\noindent Hence, for $\gamma>\mu_0$, asymptotically the brane will remain
anisotropic ($\sigma^2\sim 4\Lambda$) and expand as $a\sim t^{2/3\mu_0}$ or
expand as $a\sim t^{2/3\gamma}$ for $\gamma<\mu_0$. Note that, the exact solution
given by (\ref{eq:part1}) and (\ref{eq:part2}) for a model with no fluid, is
also the approximated late time solution when $\mu_0=4/3\le\gamma$
or $\gamma=4/3\le\mu_0$. In both cases asymptotically the scale factor behaves a
$a\sim t^{1/2}$.

In (\ref{sp}) we assumed that $\sigma^{\mu\nu}P_{\mu\nu}$ depends on $a$ and $\mathcal U$.
We could instead assume that it depends only on $a$ so 
that asymptotically (\ref{eq:assumption})
holds with
 \begin{equation}\label{eq:as1}
 \Gamma(a)=\Lambda+\frac{\beta}{a^{3\delta}},
 \end{equation}
for some constants $\beta$ and $\delta>0$.
It is easy to see that in this case also the asymptotic behavior of $a$ is a power law. 
For instance, if we additionally assume that $\gamma>4/3$ (or that 
there is no fluid), asymptotically the brane will remain anisotropic 
($\sigma^2\sim 4\Lambda$) and expand as $a\sim t^{2/3\delta}$ for 
$\delta<4/3$ and as $a\sim t^{1/2}$ for $\delta\ge 4/3$ (or  $\beta=0$).
For $\gamma<4/3$, one would have $a\sim t^{2/3\nu}$, with 
$\nu=\min(\gamma,\delta)$.  

%%%%%%%%%%%%%%%%%%%%%%%%
\section{Conclusions}

By making the more general assumption (\ref{eq:assumption}) on the
(unknown) influence of the bulk on the  brane, we have shown that some
conclusions on the asymptotic behavior of Bianchi I brane worlds in
Refs.~\cite{Maartens,Santos} can be generalized. Due to the nonlocal
stresses, in most of our models, the nonlocal energy does not vanish in the
limit $a\to\infty$, and the brane does not isotropize. We have
also found that, although nearly all our models inflate, there also
exist the possibility that the inflation due to the cosmological
constant might be prevented by the interaction with the bulk. Finally,
we have shown that the problem for the mean radius $a$ (as well as for
$\sigma^2$ and $\mathcal{U}$) can be completely solved in our models,
which include as particular cases the Bianchi I branes for which the
orbits and stability were analyzed in~\cite{Santos}.

%The question of whether inflation could be avoided by the interaction
%with the bulk, either through the mechanism discussed here or perhaps a
%more suitable one, rests unsolved unfortunately, for the bulk dynamics
%is unknown. 

%=========================================================================
\section*{ACKNOWLEDGMENTS}
%=========================================================================

%\begin{acknowledgments}

This work was supported by the University of Buenos Aires under Project
X223,  the Spanish Ministry of Science and Technology
jointly with FEDER funds through research grant  BFM2001-0988,
and the University of the Basque Country through research grant 
UPV00172.310-14456/2002. Ruth Lazkoz's
work is also supported by the Basque Government through fellowship BFI01.412.

%\end{acknowledgments}

%==================================================================


\begin{thebibliography}{99}
%==================================================================
\bibitem{Maartens}R.\ Maartens, V.\ Sahni, and T.\ D.\ Saini, Phys.\ Rev.\ D \textbf{63},
063509 (2001).

\bibitem{Santos}M.\ G.\ Santos, F.\ Vernizzi, and P.\ G.\ Ferreira, Phys.\ Rev.\ D \textbf{64},
063506 (2001).

\bibitem{Horava}P.\ Horava and E. Witten, Nucl.\ Phys.\ B \textbf{460}, 
506 (1996).

\bibitem{Randall} L.\ Randall, R.\ Sundrum, Phys. Rev. Lett. \textbf{83}, 4690 (1999).

\bibitem{Shiromizu}T.\ Shiromizu, K.\ Maeda, and M.\ Sasaki, Phys.\ Rev.\ D \textbf{62}, 
024012 (2000).

\bibitem{Maartens-rev} R.\ Maartens,
Phys. Rev. D \textbf{62}, 084023 (2000).

\bibitem{Langlois} D.\ Langlois, Astrophys Space Sci. \textbf{283},  469 (2003).

\bibitem{Brax}  Ph.\ Brax and C. van de Bruck, Class. Quant. Grav. {\bf 20},  R201 (2003)

\bibitem{Binetruy}P.\ Bin\'etruy, C.\ Deffayet, and  D.\ Langlois, Nucl.\ Phys.\ B \textbf{565}, 269 (2000).

\bibitem{Campos} A. Campos and C.\ F.\ Sopuerta,  Phys.\ Rev.\ D \textbf{63}, 104012 (2001).

\bibitem{Wald} R.\ Wald, Phys.\ Rev.\ D \textbf{28}, 2118 (1982).

\bibitem{Paul}B.\ C.\ Paul, Phys.\ Rev.\ D \textbf{66}, 124019 (2002).

\bibitem{Hoogen} R. J. van den  Hoogen, A.A. Coley. Y. He, Phys. Rev. D {\bf 68}  023502 (2003).

\bibitem{Barrow} J.D. Barrow and R. Maartens, Phys. Lett. {\bf B 532} 153 (2002). 


\bibitem{Matravers}A.\ Campos, R.\ Maartens, D. Matravers, and C.F. Sopuerta, Phys. Rev. D {\bf 68}, 103520 (2003).

\bibitem{Toporensky}A.\ V.\ Toporensky, Class.\ Quan.\ Grav. \textbf{18}, 2311 (2001).

\bibitem{Sopuerta} A. Campos and C.F. Sopuerta,  Phys. Rev. D {\bf 64}, 104011 (2001).

\bibitem{Savchenko} N. Yu. Savchenko and V.A. Toporensky, Class. Quant. Grav. {\bf 20},  2553 (2003).
 



\bibitem{Dabrowski}M.P. Dabrowski, W. Godlowski, and  M. Szydlowski, {\tt astro-ph/0212100}.

\bibitem{Heckmann} O. Heckmann and E. Schucking, Handb. Phys. {\bf 53}, 489 (1959).

\bibitem{Khalatnikov} I.\ M.\ Khalatnikov and A.\ Yu.\ Kamenshchik,
Phys. Lett. B {\bf 553}, 119 (2003).



\end{thebibliography}
\end{document}